\documentclass[10pt,journal,compsoc]{IEEEtran}

\usepackage{amsmath,amssymb}
\usepackage{booktabs}
\usepackage{tabularx}
\usepackage{array}
\usepackage{graphicx}
\usepackage{cite}
\usepackage{url}
\usepackage{microtype}
\usepackage{stfloats}

\ifdefined\pdfinfoomitdate\pdfinfoomitdate=1\fi
\ifdefined\pdftrailerid\pdftrailerid{}\fi
\ifdefined
\fi

\graphicspath{{figures/}}
\newcommand{\FprUnionIX}{0.125}
\newcommand{\FprLegacyIX}{0.061}
\newcommand{\FprFamilyIX}{0.056}

\newcommand{\FprUnionIXF}{0.203}
\newcommand{\FprLegacyIXF}{0.073}
\newcommand{\FprFamilyIXF}{0.058}

\newcommand{\FprUnionIYm}{0.048}
\newcommand{\FprLegacyIYm}{0.048}
\newcommand{\FprFamilyIYm}{0.048}

\newcommand{\FprUnionIXFY}{0.259}
\newcommand{\FprLegacyIXFY}{0.079}
\newcommand{\FprFamilyIXFY}{0.053}

\newcommand{\ConformalLevel}{0.0498}
\newcommand{\TrustedGrossExactBlocks}{85}
\newcommand{\TrustedExactOverlap}{46}
\newcommand{\TrustedNetAdditional}{39}
\newcommand{\TrustedNetAdditionalPct}{3.25}
\newcommand{\UnsafeUnionIX}{4,429}

\newcommand{\DecFprUnionIX}{0.125}

\newcommand{\UnsafeFamilyIX}{4,496}

\newcommand{\DecFprFamilyIX}{0.056}

\newcommand{\UnsafeLegacyFamilyIX}{4,494}

\newcommand{\UnsafeUnionIXF}{4,231}
\newcommand{\ContainUnionIXF}{0.40}
\newcommand{\DecFprUnionIXF}{0.203}
\newcommand{\ContainFiveUnionIXF}{0.70}

\newcommand{\UnsafeFamilyIXF}{4,365}
\newcommand{\ContainFamilyIXF}{0.38}
\newcommand{\DecFprFamilyIXF}{0.058}
\newcommand{\ContainFiveFamilyIXF}{0.66}

\newcommand{\UnsafeLegacyFamilyIXF}{4,327}

\newcommand{\UnsafeFamilyIYm}{7,008}

\newcommand{\UnsafeUnionIXFY}{4,180}
\newcommand{\ContainUnionIXFY}{0.40}
\newcommand{\DecFprUnionIXFY}{0.259}

\newcommand{\BenignHBUnionIXFY}{0.76}

\newcommand{\UnsafeFamilyIXFY}{4,390}
\newcommand{\ContainFamilyIXFY}{0.37}
\newcommand{\DecFprFamilyIXFY}{0.053}

\newcommand{\BenignHBFamilyIXFY}{0.49}

\newcommand{\BenignInterruptFamilyIXFY}{590}

\newcommand{\UnsafeLegacyFamilyIXFY}{4,322}

\newcommand{\ContainFamilyIXFYtrusted}{1.00}

\newcommand{\BenignHoldFamilyIXFYtrusted}{544}

\newcommand{\BenignInterruptFamilyIXFYtrusted}{629}

\newcommand{\NMaterial}{7,008}

\newcommand{\ResidualBlindFamilyIYm}{5,450}
\newcommand{\NMaterialLabel}{2,617}
\newcommand{\BatchUnionLabelFires}{43}

\newcommand{\BatchFamilyLabelFires}{11}

\newcommand{\WitnessWOne}{808}
\newcommand{\WitnessWOneDen}{3,600}

\newcommand{\WitnessWSeven}{1,534}

\newcommand{\WitnessWEight}{2,617}
\newcommand{\WitnessWEightDen}{2,617}
\newcommand{\WitnessWNine}{2,792}

\newcommand{\WitnessWTen}{808}

\newcommand{\NMaterialFamDropout}{1,079}

\newcommand{\UnsafeFamilyDropoutIXF}{1,057}

\newcommand{\UnsafeFamilyLabelIXFY}{2,606}

\newcommand{\NEvalDraws}{12,000}

\newcommand{\LabelDecreasedExp}{2,184}
\newcommand{\LabelUnchangedExp}{983}
\newcommand{\LabelIncreasedExp}{433}
\newcommand{\LabelMaterialExp}{2,617}
\newcommand{\LabelRowsExp}{3,600}
\newcommand{\LabelDecreasedGone}{1,276}
\newcommand{\LabelUnchangedGone}{105}
\newcommand{\LabelIncreasedGone}{59}
\newcommand{\LabelMaterialGone}{1,335}
\newcommand{\LabelRowsGone}{1,440}

\newcommand{\AdvDetAdaptSDTenPctIX}{0.56}
\newcommand{\AdvDetAdaptSDTenPctIXF}{0.66}

\newcommand{\AdvDetCtrlIXMin}{0.99}
\newcommand{\AdvDetCtrlIXMax}{1.00}

\newcommand{\AdvDetCtrlHeadlineMin}{0.96}
\newcommand{\AdvDetCtrlHeadlineMax}{1.00}

\newcommand{\AdvDetAdaptHeadlineMin}{0.01}
\newcommand{\AdvDetAdaptHeadlineMax}{0.66}

\newcommand{\AdvMatRetentionMinPct}{83}
\newcommand{\AdvMatRetentionMaxPct}{91}

\newcommand{\GeoIdentityMax}{0}
\newcommand{\GeoCleanFamilyIX}{0.056}
\newcommand{\GeoCleanFamilyIXF}{0.058}

\newcommand{\GeoTinyGaussianIXF}{0.568}
\newcommand{\GeoTinyScalingIX}{0.822}

\newcommand{\GeoDropoutFamilyMin}{0.068}
\newcommand{\GeoDropoutFamilyMax}{0.320}
\newcommand{\GeoGateLowerCellsMin}{25}
\newcommand{\GeoGateLowerCellsMax}{40}
\newcommand{\GeoGateCellTotal}{40}
\newcommand{\GeoGateAdaptiveMaterialMin}{0.262}
\newcommand{\GeoGateAdaptiveMaterialMax}{0.667}

\renewcommand{\FprUnionIX}{0.123}
\renewcommand{\FprLegacyIX}{0.063}
\renewcommand{\FprFamilyIX}{0.056}

\renewcommand{\FprUnionIXF}{0.201}
\renewcommand{\FprLegacyIXF}{0.074}
\renewcommand{\FprFamilyIXF}{0.059}

\renewcommand{\FprUnionIYm}{0.048}
\renewcommand{\FprLegacyIYm}{0.048}
\renewcommand{\FprFamilyIYm}{0.048}

\renewcommand{\FprUnionIXFY}{0.257}
\renewcommand{\FprLegacyIXFY}{0.080}
\renewcommand{\FprFamilyIXFY}{0.055}

\renewcommand{\UnsafeUnionIX}{4,429}

\renewcommand{\DecFprUnionIX}{0.123}

\renewcommand{\UnsafeFamilyIX}{4,496}

\renewcommand{\DecFprFamilyIX}{0.056}

\renewcommand{\UnsafeLegacyFamilyIX}{4,490}

\renewcommand{\UnsafeUnionIXF}{4,233}
\renewcommand{\ContainUnionIXF}{0.40}
\renewcommand{\DecFprUnionIXF}{0.201}

\renewcommand{\ContainFiveUnionIXF}{0.70}
\renewcommand{\UnsafeFamilyIXF}{4,368}
\renewcommand{\ContainFamilyIXF}{0.38}
\renewcommand{\DecFprFamilyIXF}{0.059}

\renewcommand{\ContainFiveFamilyIXF}{0.66}

\renewcommand{\UnsafeLegacyFamilyIXF}{4,329}

\renewcommand{\UnsafeFamilyIYm}{7,008}

\renewcommand{\UnsafeUnionIXFY}{4,182}
\renewcommand{\ContainUnionIXFY}{0.40}
\renewcommand{\DecFprUnionIXFY}{0.257}
\renewcommand{\BenignHBUnionIXFY}{0.75}

\renewcommand{\UnsafeFamilyIXFY}{4,394}
\renewcommand{\ContainFamilyIXFY}{0.37}
\renewcommand{\DecFprFamilyIXFY}{0.055}
\renewcommand{\BenignHBFamilyIXFY}{0.48}

\renewcommand{\BenignInterruptFamilyIXFY}{576}

\renewcommand{\UnsafeLegacyFamilyIXFY}{4,325}

\renewcommand{\ContainFamilyIXFYtrusted}{1.00}

\renewcommand{\BenignHoldFamilyIXFYtrusted}{532}

\renewcommand{\BenignInterruptFamilyIXFYtrusted}{617}

\renewcommand{\TrustedGrossExactBlocks}{85}
\renewcommand{\TrustedExactOverlap}{44}
\renewcommand{\TrustedNetAdditional}{41}
\renewcommand{\TrustedNetAdditionalPct}{3.42}

\renewcommand{\UnsafeFamilyDropoutIXF}{1,058}

\renewcommand{\UnsafeFamilyLabelIXFY}{2,606}

\renewcommand{\AdvDetAdaptSDTenPctIX}{0.57}
\renewcommand{\AdvDetAdaptSDTenPctIXF}{0.65}

\renewcommand{\AdvDetCtrlIXMin}{0.99}
\renewcommand{\AdvDetCtrlIXMax}{1.00}

\renewcommand{\AdvDetCtrlHeadlineMin}{0.94}
\renewcommand{\AdvDetCtrlHeadlineMax}{1.00}

\renewcommand{\AdvDetAdaptHeadlineMin}{0.01}
\renewcommand{\AdvDetAdaptHeadlineMax}{0.65}

\renewcommand{\GeoIdentityMax}{0}
\renewcommand{\GeoCleanFamilyIX}{0.056}
\renewcommand{\GeoCleanFamilyIXF}{0.059}

\renewcommand{\GeoTinyGaussianIXF}{0.540}
\renewcommand{\GeoTinyScalingIX}{0.723}

\renewcommand{\GeoDropoutFamilyMin}{0.063}
\renewcommand{\GeoDropoutFamilyMax}{0.303}
\renewcommand{\GeoGateLowerCellsMin}{25}
\renewcommand{\GeoGateLowerCellsMax}{40}
\renewcommand{\GeoGateCellTotal}{40}
\renewcommand{\GeoGateAdaptiveMaterialMin}{0.262}
\renewcommand{\GeoGateAdaptiveMaterialMax}{0.667}

\newcommand{\AlignedLabelFamilyFires}{343}
\newcommand{\AlignedLabelMaterialDen}{2,700}
\newcommand{\AlignedLabelUnionFires}{1,183}
\newcommand{\AlignedLabelSeparableFamilyFires}{352}
\newcommand{\AlignedLabelSeparableUnionFires}{1,204}
\newcommand{\AlignedLabelSeparableDen}{2,836}
\newcommand{\AlignedLabelBlindDen}{764}

\newcolumntype{Y}{>{\raggedright\arraybackslash}X}

\newcommand{\IX}{\mathcal I_X}
\newcommand{\IXF}{\mathcal I_{XF}}
\newcommand{\IYm}{\mathcal I_{Y_m}}
\newcommand{\IXFY}{\mathcal I_{XFY}}
\newcommand{\IXFYs}{\mathcal I_{XFY}^{\star}}
\newcommand{\IQ}{\mathcal I_Q}
\title{Observational Indistinguishability and Integrity Blind Regions in Hybrid Quantum-Classical Workflows}

\author{Roberto~Fern\'andez-Barrios,
        Iker~Pastor-L\'opez,
        Amaia~Pikatza-Huerga,
        and~Pablo~Garc\'ia~Bringas%
\IEEEcompsocitemizethanks{%
\IEEEcompsocthanksitem The authors are with the Faculty of Engineering,
University of Deusto, Avda.\ de las Universidades 24, 48007 Bilbao, Spain.
E-mail: roberto.fernandez.b@deusto.es (corresponding author),
iker.pastor@deusto.es, a.pikatza@deusto.es, pablo.garcia.bringas@deusto.es.
\IEEEcompsocthanksitem This work is part of grant PID2024-155693NB-C43,
ATHENA-AEGIS (Advanced Secure Technologies for Hybrid Quantum-Classical
Environments and Applications), funded by MICIU/AEI/10.13039/501100011033 and
by ERDF/EU.}%
}

\begin{document}

\maketitle

\begin{abstract}
We present a claim-relative evidence/reference framework for hybrid
quantum-classical workflow integrity. Observational indistinguishability yields
structural blind regions, distinct from finite-batch statistical misses. Within
the declared lattice, a trusted same-batch scalar $R_0$ suffices for conclusion
integrity, aggregate $M_0$ for aggregate plus conclusion integrity, and
item-aligned binding for item identity. In 3,600 label interventions,
feature/prediction views realize exact label-path invariance; all
\AlignedLabelBlindDen{} geometry-aligned aggregate-blind rows equal their paired-clean
responses, giving zero attack-only increment. For statistical response, the
geometry-aligned construction detects \AlignedLabelFamilyFires{}/\AlignedLabelMaterialDen{}
conclusion-changing ($\tau\to0^{+}$) label interventions with the conformal rule and
\AlignedLabelUnionFires{}/\AlignedLabelMaterialDen{} with the uncorrected union;
the original frozen same-item geometry yields 11/2,617 and 43/2,617,
respectively. The executed conformal clean false-action rates are
\FprFamilyIYm--\FprFamilyIXF{} descriptively; its finite-sample guarantee requires
exchangeability, which the overlapping-draw design violates. The
cluster-preserving adaptive stress test (Gate~A) reduces response versus matched
controls in 25--40 of 40 environment/split cells while retaining conclusion
changes. A bounded
165-design-cell ideal-statevector and finite-shot-emulation branch directly instantiates
semantic, estimated and observed kernel transitions. The fixed equal-weight
design estimates neither deployment prevalence nor QPU, provider or deployed-service assurance.
\end{abstract}


\section{Introduction}

\IEEEPARstart{H}{ybrid} quantum-classical software distributes one reported
result across heterogeneous computational and evidential boundaries. Classical
records are acquired, cleaned, projected and encoded; a circuit and execution
stack produce quantum evidence; a learner turns a kernel into a decision; and
an evaluation layer aligns that decision with reference outcomes and reports a
metric. A plausible final number may therefore coexist with a failure in the
data, circuit, kernel, prediction, label or reporting path. Security analyses
of hybrid systems consequently call for lifecycle-aware controls rather than
trust in an isolated algorithm \cite{Volya2023,Ghosh2025}.

Robustness benchmarking collapses this chain into a performance change. That
endpoint cannot identify what changed or whether the auditor could have seen
it. Consider a hybrid intrusion-detection service that scores each batch with a fixed
predictor, then joins the predictions to a ground-truth feed from another
subsystem (analyst verdicts, incident tickets or a delayed label store). The
predictor is intact, but the join or store is stale, corrupted or rewritten,
so balanced accuracy moves although the model has not. Feature and prediction
monitors cannot see the event because their inputs are unchanged. Label counts
expose an unbalanced corruption but not a balanced exchange of class identities.
A signature authenticates the label file, not its item-level alignment with the
predictions, so a correctly signed stale or misaligned file still passes.

Calling the event stealthy therefore requires a qualifier: stealthy to which
auditor, holding which evidence and trusting which reference? The constructive
result is claim-relative. A trusted scalar reference $R_0$ for the same batch
certifies whether the reported conclusion changed. A trusted aggregate
$M_0$ additionally certifies aggregate integrity and, by Proposition~3, every
material conclusion change in the declared setting; only item-identity
integrity requires item alignment. If one party
controls both the label store and its checking reference, no local verifier can
distinguish a substituted honest pair from the original (Proposition~6). The
paper identifies a necessary external root and claim-relative sufficient
granularity within the declared reference model; it does not remove the need
for that root. The scenario is a
reading aid, not an incidence estimate.

Prior art establishes observation-relative detectability
\cite{Pasqualetti2013,Teixeira2015,Liu2011}, authenticated evaluation artifacts
\cite{Stokes2021}, detector-aware drift \cite{Hinder2025} and quantum-stage
contracts \cite{YeniarasKarimov2026,Yeniaras2026PipeGuard}. We claim none of
these building blocks individually. Their composition here yields a
claim-relative integrity lattice with structural blind regions, an explicit
evaluation-label path, reference sufficiency/necessity for three claim levels,
and a separation between structural blindness and finite-sample miss. The
experiments instantiate those distinctions, policy consequences, one declared
adaptive fingerprint and a bounded quantum branch.

The contribution is one package:
\begin{itemize}
  \item a claim-relative workflow view/reference lattice, with monotonicity,
  closure, materiality and minimal counterexamples, including distinct
  semantic, estimated and observed kernels;
  \item an explicit adversary/failure model, including an adaptive attacker
  that preserves the declared batch fingerprint;
  \item multi-dataset fixed-OOD validation with duplicate-controlled result
  aggregation, numerical identity checks, cluster-level uncertainty and
  prespecified null calibration;
  \item a conformal family rule valid under exchangeability and an offline
  \texttt{allow/hold/block} analysis over the frozen grid; and
  \item a bounded quantum-path instantiation and four-contract prototype.
\end{itemize}

The lattice and its claim-relative derivation are the contribution; Max-Rank,
contracts, hash chains and cryptographic roots remain prior-art components.
Results have three layers: structural propositions, finite-design statistical
responses and constructive exact checks. Hardware, calibrated backend noise,
provider security, deployed services and the ATHENA fleet-management case study
(WP5) are outside scope (subsequent operational work of the ATHENA-AEGIS
subproject); the supplemental model profile is neither quantum advantage nor
vulnerability ordering.

\section{Related Work and Positioning}

\subsection{Stealth, Detectability and Observability in Secure Systems}

Observation-relative detectability is established: cyber-physical attacks can
be unobservable, residual-stealthy or confined to an estimator's range space
\cite{Pasqualetti2013,Teixeira2015,Liu2011}; later work quantifies the
detectability--impact trade-off and the residuals a monitor can form
\cite{Bai2017,Urbina2016,Giraldo2018}. Hinder \emph{et al.} construct drifts
designed to evade drift detectors \cite{Hinder2025}. Our cluster-preserving
adaptive stress test (Gate~A, Section~V-E) instantiates this against the
declared batch fingerprint; it measures conclusion-changing fractions, policy
consequences and evidence regimes without claiming priority for adaptive drift
evasion. Simplex supplies the canonical decision/fallback
pattern \cite{Seto1998,Sha2001}, runtime verification checks executions
\cite{Leucker2009}, and integrity checkers compare digests with trusted
baselines \cite{Kim1994}. Our claim is not generic information-relative
detectability, but blind regions and calibrated decisions for workflow views
whose protected boundaries include evaluation ground truth.

\subsection{Monitoring, Labels and Evaluation Integrity}

Monitoring work treats shifted priors, selective labels and proxy sufficiency
under delayed ground truth \cite{Lipton2018,Ginart2022,Koebler2025,
Solozobov2026}. Benchmark-quality work treats label error, evaluation pitfalls,
spatial/temporal sampling bias and measurement blindness
\cite{Northcutt2021,Arp2022,Pendlebury2019,Bajaj2026}. These lines primarily
model labels as shifted, delayed, noisy, erroneous or measurement-limited.
VAMP \cite{Stokes2021}, by contrast, explicitly treats evaluation artifacts as
authenticated assets that may be poisoned or substituted. Our distinction is
therefore not that evaluation data can be attacked, but that auditability and
trusted-reference granularity are derived relative to the protected claim and
the auditor's view.

VAMP protects named datasets, software, models and evaluation sets through
authenticated provenance, providing a concrete external root of the kind
required by Proposition~6. Its focus is artifact authentication; our
complementary question is which view and granularity suffice for conclusion,
aggregate or item-identity integrity, and what remains indistinguishable with
weaker evidence. A VAMP-style manifest can instantiate the references assumed
here.

In-toto, SLSA and Sigstore bind supply-chain steps, build provenance, identity
and transparency \cite{TorresArias2019,SLSA2025,Newman2022}. Remote attestation
supplies fresh evidence to an appraiser \cite{Coker2011}; RATS names Attesters,
Verifiers, Evidence, Reference Values and appraisal policy \cite{Birkholz2023}.
These mechanisms establish provenance or appraisal, not whether aggregate or
item-aligned evidence suffices for a claim; provider/runtime authentication is
subsequent operational work of the ATHENA-AEGIS subproject.

\subsection{Calibration of Multi-Sensor Decisions}

The family rule of Section~III-D is a conformal $p$-value
\cite{Vovk2005,Angelopoulos2023}. We use the Max-Rank statistic of Timans
et al. within a full-conformal family construction \cite{Timans2025}: the
maximum of the per-sensor ranks is equivalently a Tippett minimum-$p$
ordering, closely related to Westfall--Young resampling corrections
\cite{Tippett1931,Westfall1993}. The augmented leave-one-out construction and
ties-against-rejection convention supply the stated finite-sample argument
under exchangeability \cite{Vovk2005,Lei2018}. Conformal anomaly detection and
shared-calibration $p$-values provide the surrounding statistical context
\cite{Laxhammar2015,Bates2023}. We claim no novelty for the statistic or rule:
the family score calibrates an information-regime decision under its stated
premise and exposes the executed design's premise violation. Beyond-exchangeability conformal
methods exist \cite{Barber2023}, but this paper does not implement one. The
asymmetric comparison rule had no such guarantee (Proposition~5(c)).

\subsection{Hybrid and Quantum Software Integrity}

Hybrid-system analyses decompose control, compilation, execution and
measurement surfaces and cover malicious compilation, hardware faults, cloud
exposure and side channels \cite{Volya2023,Ghosh2025,Saki2022}. Quantum
Vulnerability Factor scores fault sensitivity, while Quantum Leak demonstrates
cloud timing leakage \cite{Oliveira2024,Lu2025}. QCIVET combines stage
contracts, hash-chained traces, behavioural subtyping, calibrated observable
tests and QPU validation \cite{YeniarasKarimov2026}. QML-PipeGuard uses a
structured observable family and calibrated drift tolerance to detect channel
substitution on IBM hardware \cite{Yeniaras2026PipeGuard}; multi-level work
evaluates circuit-integrity metrics \cite{Ahmed2026}.

Bensoussan \emph{et al.} derive a taxonomy from real hybrid-software faults
\cite{Bensoussan2026Taxonomy}; our synthetic interventions are controlled
distinguishability probes, not a prevalence model. Quantum Squeeziness
formalizes quantum-program testability and fault masking
\cite{Bensoussan2026Squeeziness}; we instead study workflow-wide observational
indistinguishability across a claim-relative evidence lattice, including
evaluation labels and trusted-reference granularity.

QProv records provider-independent provenance across circuit, compilation,
execution and hardware context \cite{Weder2021}; recent cross-provider work
extends this line toward unified quantum-job provenance \cite{Peltonen2026}.
Differential, metamorphic and equivalence-modulo-input tests address the oracle
problem \cite{Wang2021,Paltenghi2023,Luo2026}; compiler verification, program
logics and runtime assertions supply other formal/runtime mechanisms
\cite{Shi2019,Ying2011,Zhou2019,Li2020,Yamaguchi2023}. We claim no priority for
these mechanisms. QCIVET and QML-PipeGuard reason directly about quantum-stage
contracts and behavioural evidence, including calibrated observables, drift
and hardware validation; QProv and cross-provider work address provenance.
They are stronger here on QPU/hardware and runtime/provider evidence. Our
orthogonal contribution is a workflow-wide, claim-relative information-set
lattice that protects the evaluation-label path and distinguishes conclusion,
aggregate and item-identity integrity. Its quantum evidence is simulator-only,
its hash chain unsigned, and its policy offline.

\subsection{Relation to Companion Work}

Three companion studies by the authors are disclosed to delimit the claim.
\emph{Sharp target-domain certificates} \cite{FernandezBarrios2026Certificates}
identifies where a fixed candidate advantage is supported under shift.
\emph{Conditional validity} \cite{FernandezBarrios2026Conditional} studies
conditional evidence and abstention under collider and estimation uncertainty.
\emph{Candidate comparability} \cite{FernandezBarrios2026VBC} governs
challenger promotion after drift. This paper instead concerns adversarial
workflow integrity: observational blindness, trusted roots, label-store
substitution and reference granularity. Shared data sources, fidelity kernels
and selected monitors are disclosed; the formal objects, interventions,
endpoints, tables and primary experiments are distinct. Operational QPU/context
assurance belongs to subsequent operational work of the ATHENA-AEGIS subproject.

\section{Observation Model and Blind Regions}

The supplement gives complete proofs. The formal core derives which evidence
separates each intervention class and integrity claim.

\subsection{States, Interventions and Materiality}

A workflow state separates stored artifacts from those the workflow derives.
The primitive artifacts of one evaluation run are
\begin{equation}
  s=(X,P,C,E,\xi,g,f,y)\in\mathcal S,
  \label{eq:workflow}
\end{equation}
acquired features $X$, preprocessing $P$, circuit or feature map $C$,
execution context $E$, estimation randomness $\xi$, post-processing map $g$,
fitted predictor $f$ and evaluation labels $y$. The derived artifacts are
recomputed from them: the representation $\tilde X=P(X)$; the semantic kernel
$K_{\mathrm{sem}}=\Phi(C)$ induced by the circuit on a fixed probe set; the
finite-shot estimate $\hat K=\mathrm{est}(K_{\mathrm{sem}},E,\xi)$; the
observed kernel $K_{\mathrm{obs}}=g(\hat K)$ delivered downstream after
estimation, transmission and post-processing; the predictions
$\hat y=f(\tilde X,K_{\mathrm{obs}})$, indexed by item identity
$i=1,\dots,n$; and the result $R=r(\hat y,y)$, balanced accuracy of $\hat y$
against $y$. An intervention $a$ overwrites a declared set of artifacts,
primitive or derived, keeps every non-descendant fixed and recomputes the
descendants through the workflow with $\xi$ retained; the baseline is $s_0$
and $s_a=a(s_0)$. Label-only $T_y$ overwrites $y$ item-wise, so only $R$ is
recomputed (its prior-preserving subclass $T_y^{\pi}$ also preserves the
class histogram); feature-side mechanisms overwrite $\tilde X$ or $X$ and
recompute $\hat y$ and $R$; circuit-side interventions overwrite $C$ and
recompute $K_{\mathrm{sem}}$ and everything below it; estimation variation
overwrites $\xi$ or $E$ with $K_{\mathrm{sem}}$ fixed; post-processing
overwrites $g$ or $K_{\mathrm{obs}}$ with $K_{\mathrm{sem}}$ and $\hat K$
fixed. ``Fixes everything else'' thus means every non-descendant. The signed
conclusion change is
\begin{equation}
  \Delta_R(a)=R(s_0)-R(s_a).
  \label{eq:conclusion-impact}
\end{equation}
An intervention is \emph{material at level} $\tau$ if $|\Delta_R(a)|\ge\tau$.
The limit $\tau\to0^{+}$ is a structural-sensitivity endpoint---whether the
reported conclusion changes at all---not an operational risk threshold or an
assertion that every epsilon has the same consequence. The fixed
$\tau=0.02$ and $0.05$ analyses
show larger effect sizes; service-level risk is outside this study. The
signed endpoint, unlike the positive part $\max(\Delta_R,0)$, counts an
apparent improvement caused by an integrity failure as a conclusion change.

\subsection{Views, Refinement and Trusted References}

An information set $\mathcal I$ is a view map
$V_{\mathcal I}:\mathcal S\to\mathcal O_{\mathcal I}$; the auditor observes
$V_{\mathcal I}(s)$ and nothing else. The regimes studied are $\IX$
(the multiset of rows of $\tilde X$), $\IXF$ (the multiset of pairs
$(\tilde x_i,\hat y_i)$), $\IYm$ (the class-count vector of $y$), $\IXFY$ (the
multiset of triples $(\tilde x_i,\hat y_i,y_i)$) and $\IQ$ (circuit, kernel
and execution evidence). $\mathcal I\sqsubseteq\mathcal I'$ (refinement) iff
$V_{\mathcal I}=\pi\circ V_{\mathcal I'}$ for some $\pi$; hence
$\IX\sqsubseteq\IXF\sqsubseteq\IXFY$ and $\IYm\sqsubseteq\IXFY$, while $\IYm$
is incomparable with $\IX$ and $\IQ$ is a separate branch. The join
$\mathcal I\vee\mathcal W$ is the pair of views.

A \emph{reference profile} separates three dimensions: provenance
(historical, benchmark-protected, or deployment-authenticated), granularity
(aggregate or item-aligned), and decision rule (statistical or exact).
A reference is \emph{trusted} if no intervention in the class under study can
alter it. The A--C labels are shorthand, not a one-dimensional trust scale.
Class~A is a \emph{statistically thresholded aggregate comparison}; its
reference may be historical or, as executed here, a protected clean
same-item-set oracle used without item pairing. The oracle is outside the
benchmark attack API, but no deployed authentication mechanism is
demonstrated. Class~B is an exact invariant against a trusted aggregate
same-batch reference, such as $M(s_0)$, $\mathrm{hist}(y_0)$ or $R(s_0)$.
Class~C is an exact invariant against a trusted item-aligned same-batch
reference such as $(y_{0,i})_i$. Thus statistical versus exact and aggregate
versus item-aligned are not synonyms for untrusted versus trusted. For a label-path
intervention, \emph{item-identity integrity} holds if $y_a=y_0$ item-wise,
\emph{aggregate integrity} if $M(s_a)=M(s_0)$ and \emph{conclusion
integrity} if $R(s_a)=R(s_0)$; each implies the next and no converse holds.
$\mathcal I^{\star}$ denotes a regime augmented with trusted same-batch
references ($\IXFYs$ holds the aggregate confusion profile and item-aligned
labels and predictions; feature coverage remains statistical). A label
available in $\IXFY$ is not thereby trusted: if the label store is the asset
under attack, the auditor sees $y_a$, not $y_0$.

\subsection{Equivalence, Sensors and Blind Regions}

\textbf{Definition 1.} $s\sim_{\mathcal I}s'$ iff
$V_{\mathcal I}(s)=V_{\mathcal I}(s')$. A sensor admissible in $\mathcal I$ is a
measurable $S:\mathcal O_{\mathcal I}\to\mathbb R^k$, written
$S(s)=S(V_{\mathcal I}(s))$.

\textbf{Lemma 1 (structural blindness).} If $s_a\sim_{\mathcal I}s_0$ then
$S(s_a)=S(s_0)$ for every sensor admissible in $\mathcal I$ (path-wise for a
seeded randomized sensor; in distribution for a fresh seed).

\textbf{Definition 2.} For a class $\mathcal A$, the structural blind region
of $\mathcal I$ is
$B_{\mathcal I}(\mathcal A)=\{a\in\mathcal A: a(s_0)\sim_{\mathcal I}s_0\}$ and
the sensor blind region of a family $\mathcal S$ is
$B_{\mathcal I,\mathcal S}(\mathcal A)=\{a:S(a(s_0))=S(s_0)\ \forall S\in\mathcal S\}$.
By Lemma~1, $B_{\mathcal I}\subseteq B_{\mathcal I,\mathcal S}$, with equality
iff $\mathcal S$ is \emph{baseline-separating} on the orbit of $\mathcal A$
(a different value at $a(s_0)$ than at $s_0$ whenever the views differ);
detection needs baseline separation, identification would need pairwise
separation, which is not claimed. Sensor blindness thus decomposes into
information-level blindness and sensor insufficiency; the adaptive attacker
of Section~VI-D exploits the second. The structural and sensor auditability
gaps at level $\tau$ are
\begin{equation}
  G_{\mathcal I}(\tau)=\{a:|\Delta_R(a)|\ge\tau,\ a\in B_{\mathcal I}\},\quad
  G_{\mathcal I,\mathcal S}(\tau)\supseteq G_{\mathcal I}(\tau).
  \label{eq:gap}
\end{equation}

\textbf{Proposition 1 (monotonicity).} If $\mathcal I\sqsubseteq\mathcal I'$
then $B_{\mathcal I'}(\mathcal A)\subseteq B_{\mathcal I}(\mathcal A)$ and
$G_{\mathcal I'}(\tau)\subseteq G_{\mathcal I}(\tau)$ for every $\tau$.

\begin{IEEEproof}
$V_{\mathcal I'}(s_a)=V_{\mathcal I'}(s_0)$ implies
$V_{\mathcal I}(s_a)=\pi(V_{\mathcal I'}(s_a))=\pi(V_{\mathcal I'}(s_0))=V_{\mathcal I}(s_0)$.
\end{IEEEproof}

\textbf{Corollary 1 (label-path boundaries).} Let $f$ be fixed and
deterministic. (a) $T_y\subseteq B_{\IXF}\subseteq B_{\IX}$: every sensor
admissible in $\IX$ or $\IXF$ is invariant under a label-only change.
(b) $T_y^{\pi}\subseteq B_{\IYm}$: every sensor admissible in $\IYm$ is
invariant under a prior-preserving change.

\textbf{Proposition 2 (closure).}
$B_{\mathcal I\vee\mathcal W}(\mathcal A)=B_{\mathcal I}(\mathcal A)\cap B_{\mathcal W}(\mathcal A)$;
a class $\mathcal A\subseteq B_{\mathcal I}$ becomes completely separable
after adding $\mathcal W$ iff $V_{\mathcal W}(a(s_0))\ne V_{\mathcal W}(s_0)$
for every $a\in\mathcal A$.

\textbf{Corollary 2 (label-path closure).} No view factoring through
$(\tilde X,\hat y,\mathrm{hist}(y))$ separates any $a\in T_y^{\pi}$.
The multiset $\IXFY$ separates it unless labels move only among items with
identical $(\tilde x,\hat y)$; an item-aligned view separates every
non-identity relabeling.

\textbf{Proposition 3 (materiality forces aggregate separability).} Let
$R=g(M)$ be a function of the confusion matrix $M(s)$, as balanced accuracy
is. If $a\in T_y$ and $\Delta_R(a)\ne0$ then $M(s_a)\ne M(s_0)$; hence $a$ is
separable in the confusion view and in $\IXFY$, and
$G_{\IXFY}(\tau)\cap T_y=\emptyset$ for every $\tau>0$.

\begin{IEEEproof}
Contrapositive of $M(s_a)=M(s_0)\Rightarrow R(s_a)=R(s_0)$; the confusion view
is a coarsening of the multiset of triples.
\end{IEEEproof}

The proposition holds for every metric that is a function of the binarised
confusion matrix; for a ranking metric such as ROC-AUC the same argument holds
with the multiset of (score, label) pairs as the aggregate (supplement). No
AUC experiment is run.

\subsection{Three Auditor Classes and Decision-Level Calibration}

A \emph{reference-anchored} auditor holds a trusted same-batch reference
$\rho$ (class B or C) and computes $D(s)=d(V_{\mathcal J}(s),\rho)$ with a
metric $d$; then $D(s_0)=0$ exactly, the rule ``fire iff $D>0$'' has zero
false-action probability, and separability in $\mathcal J$ is sufficient for
detection. A \emph{batch-level statistical} auditor computes aggregate
scores and thresholds them against a clean calibration population (class A).
Its scoring reference can be historical or same-item-set; either way,
separability need not yield power beyond clean-score variability. A
\emph{post-hoc certifier}
recomputes a view from inputs it trusts (the evaluation contract recomputes
$R$ from $(\hat y,y)$) and needs trusted inputs rather than a stored value.

\textbf{Proposition 4.} (i) For any auditor whose decision is a function of
$V_{\mathcal I}(s)$, $a\in B_{\mathcal I}$ implies
$\mathrm{fire}(s_a)=\mathrm{fire}(s_0)$ pathwise under the same state and
reference construction. This equality does not in general imply equality with
a false-action rate estimated under a different clean-resampling or reference
construction. (ii) A reference-anchored auditor
detects every $a\notin B_{\mathcal J}$, deterministically.

\textbf{Corollary 3 (which reference certifies which integrity).} Let
$a\in T_y$ with $f$ fixed. (a) A trusted exact same-batch claim reference
$R_0=R(s_0)$ detects exactly every $R(s_a)\ne R_0$ and suffices for a
conclusion-only claim. A trusted $M_0=M(s_0)$ detects every violation of aggregate integrity
and, by Proposition~3, every material conclusion change. (b) References that
factor through $M$, $\mathrm{hist}(y)$ or $R$ miss C1 relabelings; item identity
needs level C, whose aligned $y_0$ detects every non-identity relabeling. Thus
$R_0$, $M_0$ and item alignment protect conclusion, aggregate-plus-conclusion
and identity claims, respectively. Their sufficiency and the counterexample
necessity are claim-relative within the declared reference lattice; the
minimality statement is not over every encoding or audit architecture.

\textbf{Proposition 5 (union versus conformal family calibration).} (a) If
$m$ sensors fire under the null with probabilities $p_j\le\alpha$, then
$\max_j p_j\le\Pr_0[\bigcup_j E_j]\le\min(1,\sum_j p_j)\le\min(1,m\alpha)$,
the bounds being attained by nested and disjoint events. (b) Let $v_1,\dots,v_n$ be the sensor vectors of $n$
calibration draws and $v_{n+1}$ that of the audited batch; give every member
$j$ of the augmented set the leave-one-out family score
$\tilde U_j=\max_{s}\#\{i\ne j: v_{s,i}<v_{s,j}\}/n$, let
$p=(1+\#\{i\le n:\tilde U_i\ge\tilde U_{n+1}\})/(n+1)$, and fire iff
$p\le\alpha$ (ties against firing). At most $\lfloor\alpha(n+1)\rfloor$
members of any augmented set would fire if audited; hence, if the $n+1$ draws
are exchangeable, $\Pr_0[p\le\alpha]\le\lfloor\alpha(n+1)\rfloor/(n+1)\le\alpha$
with no continuity or tie assumption \cite{Vovk2005,Tippett1931,Westfall1993};
$10/201=\ConformalLevel{}$ for $n=200$, $\alpha=0.05$. For one sensor without
ties the rule is the per-sensor null-calibration rule (Gate~N, Section~V-C).
The level is marginal over
the joint draw of calibration set and audited batch; nothing is claimed when
the premise fails. (c) An asymmetric construction that
scored calibration draws against the other $n-1$ draws only, has no such
guarantee: with three sensors and cyclic orderings over five vectors it fires
with probability $0.60$ at $\alpha=0.2$ under exchangeability (supplement),
and on the frozen draws it exceeds $\alpha$ under exchangeable re-splits where
the conformal rule does not (Section~VI-B).

\textbf{Proposition 6 (no local authentication without an uncontrolled
root).} Let a local verifier be any map $\mathrm{acc}(V_{\mathcal I}(s),\rho)$
of the observed view and a reference bundle, with honest references
$\rho_H(s)=V_{\mathcal J}(s)$. If it accepts every honest pair and the class
can produce $(V_{\mathcal I}(s'),\rho_H(s'))$ for some honest
$s'\not\sim_{\mathcal I}s_0$ (joint control), the substitution is accepted as
an honest run of $s'$: no function of that input establishes authenticity
relative to $s_0$, and if $R(s')\ne R(s_0)$ a material change is served. A
bundle component $V_{\mathcal K}(s_0)$ the class cannot write rejects every
substitution with $V_{\mathcal K}(s')\ne V_{\mathcal K}(s_0)$
(Proposition~4(ii)): authenticity requires a root outside the declared class.
The result concerns authenticity, not consistency; by Proposition~3 a material
substitution does change the joint view.

\subsection{Counterexamples}

Eight minimal constructions separate the events a robustness number
conflates; the supplement tabulates them with the frozen expansion
observations that realise each. C1 swaps the labels of two items with equal
predictions: $M$, the histogram and $R$ are invariant while item identity
changes (\WitnessWOne{} of \WitnessWOneDen{} label rows), an integrity
violation with no conclusion impact, visible only item-wise against a trusted
reference. C2--C4 separate marginal invariance, confusion-matrix invariance
and conclusion invariance, C5--C6 do the same on the feature side, C7 shows
that the clipped ``harm'' endpoint of the earlier evidence hid
\WitnessWSeven{} conclusion changes (\LabelIncreasedExp{} on the label path),
and C8 is the empirical face of Proposition~3 (\WitnessWEight{} of
\WitnessWEightDen{} material label rows change $M$). Of the 3,600 label rows,
a trusted aggregate reference of the same batch detects \WitnessWNine{}
(every material one, Corollary~3a) and the remaining \WitnessWTen{} are C1
relabelings that only an item-aligned reference exposes (Corollary~3b). A
ninth construction, the cluster-preserving drift of Section~VI-D, is not a
structural blind region but sensor insufficiency against an adaptive
attacker, measured rather than proved.

\subsection{The Quantum Branch as an Instance of the Lattice}

Let $\mathcal C$ be the circuit representations (canonical OpenQASM~3 with
bound parameters), $\Phi:\mathcal C\to\mathcal K$ the semantic map to the
ideal fidelity kernel on a fixed probe set, $K_{\mathrm{sem}}=\Phi(C)$,
$\hat K$ and $K_{\mathrm{obs}}$ the estimated and observed kernels of
Section~III-A, $h(C)$ the provenance hash, $A(\cdot)$ the algebraic
invariants and $f_K(\tilde X)$ the decision computed from a kernel. Three
intervention classes are kept apart: (a) circuit-side, overwriting $C$; (b)
estimation variation, with $C$ and $K_{\mathrm{sem}}$ fixed and $\hat K$
changing; (c) post-processing or kernel substitution, with $C$,
$K_{\mathrm{sem}}$ and $\hat K$ fixed and $K_{\mathrm{obs}}$ changing. Each
inclusion below names its class; none is claimed across classes.

\textbf{Proposition 7 (quantum lattice).} Assume $h$ collision-free on the
circuits under study. (i) [class (a)] $[C]=\Phi^{-1}(\Phi(C))$ is the
semantic class of $C$; approved transpilation and common-unitary rewrites map
$C$ into $[C]$ without fixing $h(C)$, so provenance refines the semantic
view, $B_h\subseteq B_{K_{\mathrm{sem}}}$, strictly whenever the class
contains an approved rewrite: separability in provenance is not harm, and the
auditor needs the approved class, implemented as semantic equality of probe
kernels against the trusted $K_{\mathrm{sem},0}=\Phi(C_0)$ (level B).
(ii) [class (c)] $A$ and $f_K$ coarsen the observed kernel, so
$B_{K_{\mathrm{obs}}}\subseteq B_A$ and $B_{K_{\mathrm{obs}}}\subseteq B_f$;
a PSD-preserving substitution $K'$ with $A(K')=A(K_{\mathrm{obs},0})$ lies in
$B_A\setminus B_{K_{\mathrm{obs}}}$ and, since $h(C)$ and the semantic probe
of the unchanged circuit see nothing, is closed only by an anchored
comparison of $K_{\mathrm{obs}}$ itself: the trusted semantic probe and the
trusted observed-kernel reference are different anchors. A kernel change that
crosses no decision boundary lies in $B_f\setminus B_{K_{\mathrm{obs}}}$.
(iii) [class (b)] For an honest finite-shot estimate, exact equality
$\hat K=K_{\mathrm{sem},0}$ has false-alarm probability
$1-\Pr[\hat K=K_{\mathrm{sem},0}\mid\text{honest}]$, which depends on the
discrete support of the estimator and may be large or equal to one but is
not universally one (fidelity $1/2$ at two shots gives $1/2$). Exact equality
is therefore not an appropriate acceptance criterion;
$d(\hat K,K_{\mathrm{sem},0})$ must be treated as a level-A statistic with a
null from repeated honest estimation, to which Proposition~5(b) applies with
one sensor when the repeated and audited estimates are exchangeable, and
where no such null is calibrated no statistical integrity claim is made. The
proposition adds no quantum mechanics: it names the two features the
classical branches lack, an approved equivalence class coarser than provenance
and an estimation step that turns an exact anchor into a statistical one.
Hash chaining of the audit envelope is a reference-anchored comparison over
the record itself; its root is not assumed uncontrolled, so Proposition~6
applies.

\section{Adversary and Failure Model}

Table~\ref{tab:adversary} assigns every executed intervention to one of four
classes with the roots it cannot write and the least separating evaluated regime
(complete per-class record in the artifact). Evaluation-label classes stand
for corruption of the ground-truth store or label join; feature-side classes
for corruption or drift of the acquisition path feeding both branches; the
cluster-preserving class for an attacker who knows the batch-level
fingerprint; circuit and kernel classes for corruption of the
circuit/parameter store, the compiler output or the post-estimation kernel;
shot emulation for legitimate estimation uncertainty. Equal weighting is a
stress profile, not a threat distribution; no prevalence is estimated.

\begin{table*}[!t]
\caption{Adversary and failure model of the executed interventions. Roots: assets the class cannot write. Least separating evidence in the frozen design: the evaluated regime that first separates the class under the conformal family rule; ``exact'' means an item-aligned trusted reference; (a)--(c) are the intervention classes of Proposition~7.}
\label{tab:adversary}
\centering
\footnotesize
\setlength{\tabcolsep}{3.5pt}
\begin{tabularx}{\textwidth}{@{}p{0.24\textwidth}p{0.095\textwidth}YYp{0.235\textwidth}@{}}
\toprule
Mechanism and capability (boundary) & Class & Roots it cannot write & Least separating evidence in frozen design & Supported claim \\
\midrule
Random label flip: rewrite $r$ of $y$ at random (evaluation/report) & Integrity corruption & $X,P,C,E,\xi,g,f$; item-aligned $y^{\star}$ if declared & exact only; $\IYm$/$\IXFY$ batch 0.00--0.01 & invisible to $\IX,\IXF$ (Cor.~1a); undetected at budget by marginal or batch joint evidence at these rates \\
Prior-preserving label flip: pairwise swaps preserving class counts (evaluation/report) & Adaptive attacker & as above & exact only; $\IXFY$ batch 0.00--0.01 & invisible to $\IX,\IXF,\IYm$ (Cor.~1); every material instance changes $M$ (Prop.~3) \\
Confusion-preserving relabeling: swaps among equal predictions (realised, not executed) & Adaptive attacker & as above & item-aligned reference only & item-identity violation without aggregate or conclusion impact (C1; \WitnessWOne{} rows) \\
Feature sign flip: negate $p$ of $\tilde X$ entries (acquisition/preprocessing) & Integrity corruption & training data, fitted model, reference features & $\IX$ 0.09--0.72; $\IXF$ 0.13--0.84; exact 0.79--0.91 & partially separable; strength-dependent residual \\
Mean shift, scaling drift: shift or scale all features (acquisition/preprocessing) & Fault robustness & as above & $\IX$ \AdvDetCtrlIXMin--\AdvDetCtrlIXMax{} by cluster fingerprint & detected in this design by a fragile fingerprint \\
Feature dropout: impute $p$ of entries (preprocessing) & Fault robustness & as above & batch 0.00--0.03; exact 0.49--0.76 & separable in $\IXF$ yet missed by the executed finite-batch level-A rule \\
Cluster-preserving mean shift and scaling: same mechanisms on the 32\% of entries outside tight clusters (executed, Gate~A) & Adaptive attacker & as above & $\IX$ 0.01--\AdvDetAdaptSDTenPctIX{}, $\IXF$ 0.05--\AdvDetAdaptSDTenPctIXF{} at matched strengths; exact for every material row & reduces response against the declared fingerprint while keeping \AdvMatRetentionMinPct--\AdvMatRetentionMaxPct\% of the conclusion changes; served by P2 in 27--39\% of material rows \\
Circuit-parameter mutation; repetition change: alter $C$ before execution (design; (a)) & Adversarial attack & reference circuit hash, probe kernel $K_{\mathrm{sem},0}$ & circuit hash and semantic probe 15/15 & identified by anchored provenance and semantics; outputs miss sub-decision changes \\
Benign transpilation; common-unitary rewrite: compiler pass ((a)) & Approved equivalence & -- & hash changes 15/15, $K_{\mathrm{sem}}$ unchanged & separability in $h(C)$ is not harm (Prop.~7(i)) \\
Asymmetric, diagonal-eroded kernel edit: edit $K_{\mathrm{obs}}$ after estimation ((c)) & Integrity corruption & reference kernel, algebraic invariants & algebra 15/15 & detected by structural invariants of $K_{\mathrm{obs}}$ \\
PSD-preserving kernel substitution: valid-looking $K_{\mathrm{obs}}$ ((c)) & Adaptive attacker & reference kernel & algebra 0/15; circuit hash 0/15; anchored $K_{\mathrm{obs}}$ comparison 15/15 & sensor-level blindness closed only by a reference on the observed kernel (Prop.~7(ii)) \\
Binomial shot emulation: finite-shot estimate (execution; (b)) & Fault robustness & -- & repeated estimation 15/15 in these cells & held for adjudication; exact equality is not the criterion (Prop.~7(iii)); not hardware evidence \\
Post-hoc envelope edit: rewrite stored contracts (record) & Adversarial attack & none assumed & hash chain & tamper-evident only; no identity or non-repudiation \\
\bottomrule
\end{tabularx}
\end{table*}

The evidence trusts the local runtime and host, the training data and fitted
model, the reference input and preprocessing declaration, the reference
circuit and probe kernels, SHA-256 collision resistance, the item-level label
reference only where $\IXFYs$ is declared, and the code that builds and
verifies the envelope. Compromise of the operating system, runtime or
verifier; provider identity, forged job or calibration metadata and provider
attestation; malicious schedulers, physical mapping and unapproved passes on
hardware; device-calibrated noise, crosstalk, multi-tenant interference and
QPU faults; side channels and circuit confidentiality; denial of service; and
the ATHENA fleet-management case study (WP5) are outside this study
(subsequent operational work of the subproject).

\section{Methodology}

\subsection{Data, Pipelines and Interventions}

The frozen first gate (Gate~1) uses a balanced binary subset of CICIDS2017
\cite{Sharafaldin2018}: 3,000 rows, 77 numeric columns and 128/128 capped
training/evaluation samples. The expansion adds eight fixed environments
E1--E8 (supplement): five from CICIDS2017 (a 256/256 scale gate and four
temporal source-target pairs), two from UNSW-NB15 \cite{Moustafa2015} (a
balanced ID subset and a temporal pair) and one from ToN-IoT
\cite{Alsaedi2020} (a balanced ID subset), all at 128/128 unless stated and
projected to dimensions 8, 10 and 12. CICIDS2017 therefore carries six of
the nine environments including Gate~1; the environments are not a
probability sample of deployments. Train-defined preprocessing is applied
without evaluation labels: TruncatedSVD per branch, training-fitted
standardization for the RBF support-vector classifier (SVC) and frozen angular
scaling to $[0,2\pi]$ for the fidelity-kernel branches (ZZ, PauliXYZ and a
Z/PauliXZ-equivalent profile, one repetition). Gate~1 uses the Qiskit Machine
Learning reference evaluator \cite{JavadiAbhari2024}; the expansion uses a
tagged exact-statevector evaluator validated against it at $10^{-10}$ on unit
matrices and by an end-to-end replay (supplement); neither is shot-noise or
QPU evidence. The intervention suite has six mechanisms at strengths 0.02,
0.05 and 0.10: feature sign flip, feature-wise mean shift, scaling drift,
feature dropout with median imputation, random evaluation-label flip and
prior-preserving evaluation-label flip. Feature-only sensors are feature
Jensen--Shannon divergence (JSD), maximum mean discrepancy (MMD) and the
Kolmogorov--Smirnov (KS) rejection rate against the clean version of the same
item set; prediction, label-marginal and joint-outcome evidence likewise
compare aggregate score, class and confusion-profile distributions with
their clean same-item-set values. None uses item pairing. The runner protects
these clean arrays as an experimental oracle; the policy API receives only
sensor and family flags, but producing those flags still requires the upstream
oracle. This is statistical class A, not evidence that a deployed system
authenticates the reference. Under $\IXFYs$, confusion-profile deltas against
a trusted aggregate are exact class B invariants, while prediction
disagreement and label mismatch use trusted item alignment (class C). The
complete sensor-to-reference audit is in the supplement.

The duplicate, clustered and near-constant structure on which this benchmark
fingerprint depends is consistent with documented CICIDS2017 data-quality
limitations \cite{Engelen2021}; the frozen staging is described rather than
re-engineered here.

\subsection{Experimental Units and Uncertainty}

Gate~1 crosses five split seeds with four nested model seeds; its 7,980 raw
rows contain 3,420 exact repeated SVC rows, leaving 4,560 unique observations.
The expansion crosses the same five split seeds with two model seeds; all 360
job pairs completed and 13,680 raw rows contain 2,280 verified repeats,
leaving 11,400 unique observations. Coverage endpoints are exact counts in the
prespecified design; for the secondary model-profile endpoint (supplement) the
five split seeds are the inferential units with two-sided 95\% Student-$t$
intervals (four degrees of freedom), describing within-environment resampling
only.

\subsection{Null Calibration and Decision-Level Calibration}

The prespecified null-calibration gate (Gate~N) calibrates the ten distributional
sensors per (environment, dimension, model) cell at a nominal per-sensor
$\alpha=0.05$: the threshold is the 191st of 200 clean calibration draws from
one half of the held-out evaluation pool, and false alarms are measured on
200 draws from the disjoint other half. Gate~F (decision-level calibration)
calibrates each regime as a family with Proposition~5(b)'s conformal rule at $\alpha=0.05$ per decision,
using only the calibration draws. The same draws also evaluate the union and
asymmetric comparison rules (Proposition~5(c)). The asymmetric rule's pooled excess
over nominal is decomposed into rule bias (its rate minus the conformal rate
on identical draws) and design effect (the conformal rate minus nominal).
Both rules are also evaluated under 30 exchangeable re-splits of the 400 pooled
draws per cell. These re-splits diagnose the rule construction without changing
the nonexchangeable primary design.
Because a run's 20 draws overlap within one pool half, false-alarm
inference uses (environment, split-seed) clusters. Pooled rates are descriptive;
intervals are five-cluster $t$ intervals per environment. E1 (256-row draws
from 322-row halves) remains in the preregistered primary aggregate and is also
reported separately.

\subsection{Decision Layer and Offline End-to-End Evaluation}

Gate~D (the decision layer) composes sensors, the conformal-rule or union decision, the information regime, the
trusted-reference status, a materiality threshold and the declared residual
blind region into one \texttt{allow/hold/block} decision. \texttt{block} is
reserved for violations of an exact invariant against a trusted reference;
\texttt{hold} answers statistical evidence or, under P3, missing coverage;
the decision composes with the four frozen contracts by the maximum in
$\texttt{allow}<\texttt{hold}<\texttt{block}$. Four policies are fixed and
named by what they are: P0 serve-always, the baseline; P1, the uncalibrated
union rule of Gate~N (risk-tolerant); P2, the conformal family rule, a
\emph{conformal-rule risk-tolerant} policy, nominally calibrated under
exchangeability, that may serve under an explicitly
declared residual blind region; and P3, the \emph{sensor-coverage-complete}
policy relative to the declared evidence dimensions, which abstains (holds)
whenever a mandatory protected boundary (feature, prediction, label) has
neither exact nor declared statistical coverage under the regime. P3 fails
closed on missing coverage, not on insufficient power: a
boundary covered by a low-power statistical sensor is served as under P2, so
P3 guarantees no minimum detection power. The prototype's contracts fail
closed on their invariants; P2 does not and is never called fail-closed. Five
regimes are evaluated: $\IX$, $\IXF$, $\IYm$, $\IXFY$ (class-A statistical
aggregate comparisons) and $\IXFYs$ (class-B aggregate and class-C
item-aligned exact invariants; feature coverage remains statistical). The
policy consumes union/family/exact flags, not clean arrays; the upstream
scorer has the reference dependency described above. The observation set is
frozen: \NEvalDraws{} clean
evaluation draws, 1,200 near-null synthetic controls applied in place
(Gaussian noise $\sigma=0.001$, scaling $\alpha=0.001$; not operational
benign traffic), and the 10,800 intervened observations of the eight
environments, \NMaterial{} of which change the reported balanced accuracy
(structural-sensitivity endpoint; $\tau=0.02$ and $0.05$ are larger-effect
sensitivities, not operational risk thresholds). The primary endpoint is a
materially altered audit result served: a conclusion-changing ($\tau\to0^{+}$)
observation decided \texttt{allow}. The CSV retains \texttt{unsafe\_allow} as a compatibility
identifier; \emph{unsafe} here means serving a materially corrupted
audit/report conclusion, not allowing a malicious network event and not
necessarily changing the classifier's live prediction.
The interruption endpoints are the clean false-action rate (clean draws not allowed;
for $\IXFYs$ the denominator is the 1,200 exact-zero rows) and the
interruption rate on the near-null controls (the same 1,200 rows for every
regime), decomposed into statistical holds and exact-reference blocks. The
evaluation is offline, on frozen outputs, not a deployed runtime service;
the twelve policy-evidence consistency checks fail closed, and every table and
figure is generated from the manifested outputs. The 24,000 policy rows derived from
the frozen intervention grid are not independent episodes; inference resides
at environment/split clusters rather than at row level.

\subsection{Adversarial Gate and Geometry Sensitivity}

Gate~A executes the preregistered cluster-preserving attacker. The standardized
projected features hold
tight clusters of rows (in every environment some feature keeps 69--87\% of
the rows within 0.01 standard deviations) that a fresh clean batch reproduces
and any in-place displacement smears, which is why mean shift and scaling
drift were detected in every cell. The attacker holds the batch, knows the
sensor definitions and the fingerprint, and applies the executed mechanisms
unchanged to every unclustered entry (an entry is clustered when at least
$\lceil0.05n\rceil$ rows, never fewer than two, lie within 0.01 batch
standard deviations of it); clustered entries and labels are untouched.
Strengths 0.02, 0.05, 0.10 match the frozen suite and 0.25, 0.50 probe
whether materiality can be bought on the unclustered entries; the executed
mechanisms run in the same jobs as matched controls and must reproduce the
frozen expansion exactly. The gate runs in the eight environments with the
frozen maps, caps and seeds (240 exact-statevector jobs) and is scored with
the calibration draws, rules and policy layer of Gates~F and D; window,
minimum mass and strength grid were fixed before execution.

The geometry-aligned sensitivity leaves frozen evidence intact and replays the same 240
configurations. For fixed $E$, calibration is $s(E,C_k)$ and aligned clean
versus intervention scoring is $s(E,B)$ versus $s(E,T(B))$, with fresh $B$
fixed before results. Exact identity $s(B,B)$ is separate from clean resampling.
There is one row per frozen model cell and intervention; no label attack,
training or selection is rerun (supplement).

\subsection{Quantum Gate and Executable Prototype}

A separate simulator gate isolates the quantum boundary with CICIDS, 64/64
samples, five split seeds, dimensions 4, 6 and 8 and 11 conditions (165
cells): clean, benign level-1 transpilation and a common post-feature-map
rewrite as controls; data-dependent RZ mutations at 0.02 and 0.10; a
feature-map repetition change; asymmetric, diagonal-eroded and PSD-preserving
kernel edits; and binomial fidelity-estimation emulators at 256 and 1,024
shots. Each cell records canonical OpenQASM~3 provenance, kernel hashes, the
anchored comparison against the trusted reference kernel, algebraic checks,
repeated-estimation discrepancy and prediction changes. The prototype
instantiates four ordered contracts (input/preprocessing, circuit/kernel,
execution/result, evaluation/report) in an SHA-256 chain over one compact
CICIDS cell and six prespecified scenarios; it evaluates composition and
decision logic, not throughput. The explicit $K_{\mathrm{sem}}$, $\hat K$ and
$K_{\mathrm{obs}}$ transitions, semantic-equivalence checks and finite-shot
estimator randomness make this a quantum-stage contract path rather than a
generic ML-only pipeline. In ATHENA-AEGIS terms, the four-contract prototype
instantiates an integrity contract for the hybrid service composition exposed
as HSaaS (Hybrid Software as a Service, a composition of classical and quantum
services that the customer perceives as one service); it does not constitute
deployed HSaaS validation, and provider assurance, confidentiality, QPU
protection and the ATHENA fleet-management case study (WP5) are subsequent
operational work of the subproject.

\section{Results}

\subsection{Exact Label-Path Blind Regions}

The zero-response cells are validation checks of the structural prediction,
not findings: Corollary~1 says that a sensor whose input does not change
cannot respond, and the checks confirm that the implementation realises the
declared views. All \LabelRowsExp{} expansion and \LabelRowsGone{} Gate-1
evaluation-label observations (of 11,400 and 4,560 unique observations) have
zero prediction disagreement and exactly unchanged feature JSD, MMD, KS
rejection and score JSD; all 1,800 and 720 prior-preserving observations also
have zero prior shift and label JSD; the verifier recomputes every count. The
informative quantity is how often these blind regions coincide with material
conclusion changes and, in Section~VI-C, with serving decisions. The signed
conclusion change is not one-sided: in the expansion \LabelDecreasedExp{}
label observations lower the reported balanced accuracy, \LabelUnchangedExp{}
leave it unchanged and \LabelIncreasedExp{} raise it (Gate~1:
\LabelDecreasedGone{}, \LabelUnchangedGone{}, \LabelIncreasedGone{}). Every
one of the \LabelMaterialExp{} (Gate~1: \LabelMaterialGone{}) label
observations with a changed conclusion has non-zero item-aligned confusion
evidence, and every one of the 2,184 with a lowered conclusion, the count
used by the one-sided diagnostic, is among them.

\subsection{Decision-Level Calibration}

\begin{figure*}[!t]
  \centering
  \includegraphics[width=\textwidth]{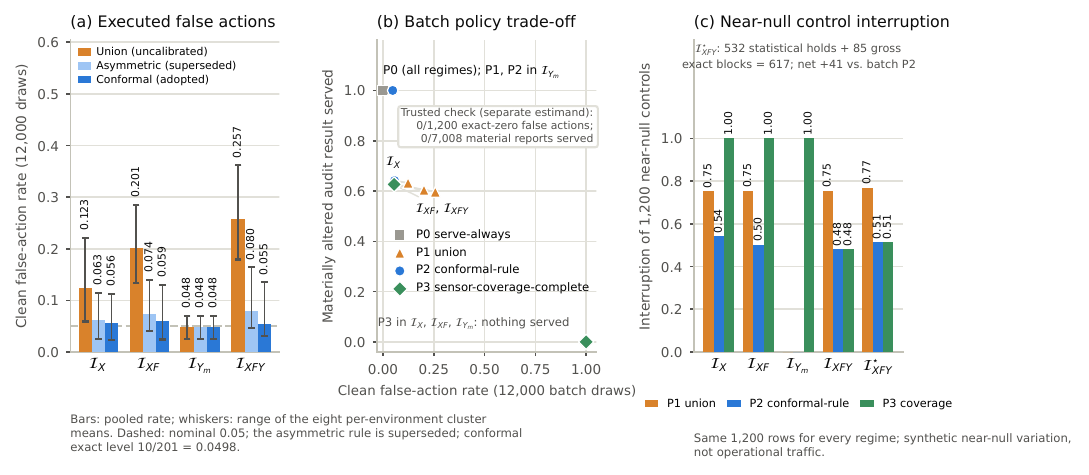}
  \caption{Gates F and D (prespecified frozen evidence). (a)
  Decision-level false-action rate on \NEvalDraws{} clean draws from evaluation
  pools disjoint from the calibration pools; resampled draws within each pool
  may overlap. Union, asymmetric-comparison and conformal rates are shown per
  regime. (b) Materially altered audit results served versus clean false-action
  rate at the structural-sensitivity endpoint (12,000-draw batch denominator);
  the trusted exact-zero check uses 1,200 rows and is a different estimand.
  (c) Interruption of 1,200 near-null controls; trusted P2 interrupts
  \BenignInterruptFamilyIXFYtrusted{}: \BenignHoldFamilyIXFYtrusted{} statistical
  holds and \TrustedGrossExactBlocks{} gross exact blocks.}
  \label{fig:policy}
\end{figure*}

Over the \NEvalDraws{} evaluation draws the per-sensor false-alarm rates lie
between 0.029 and 0.069, but the batch-level rule ``fire if any sensor
fires'' is not calibrated at the level of the decision: its rates are
\FprUnionIX{}, \FprUnionIXF{}, \FprUnionIYm{} and \FprUnionIXFY{} for $\IX$,
$\IXF$, $\IYm$ and $\IXFY$ (three, six, two perfectly dependent and ten
sensors). The conformal family rule brings them to \FprFamilyIX{},
\FprFamilyIXF{}, \FprFamilyIYm{} and \FprFamilyIXFY{} (Fig.~\ref{fig:policy}(a)),
observed rates in a design that violates the exchangeability premise. For
comparison, the asymmetric construction gives \FprLegacyIX{}, \FprLegacyIXF{},
\FprLegacyIYm{} and \FprLegacyIXFY{} on the same draws. Its excess over 0.05
contains both rule bias and design effect. Under the same exchangeable
re-splits the conformal rule stays at or below level \ConformalLevel{} in all
four regimes, while the asymmetric rule exceeds it in some.
The remaining executed-design excess is concentrated in E1; E1 stays in the
primary aggregate. Per-regime decomposition, E1/excluded-E1 values and the
unadopted split construction appear in the supplement.

The executed rules realize Proposition~4(i): across 3,600 label observations,
the $\IX/\IXF$ sensors and rules fire zero times; the $\IYm$ rules likewise fire
zero times on 1,800 prior-preserving rows. The primary empirical interpretation
of finite-batch label response uses the geometry-aligned construction
$s(E,T_y(B))$, calibration $s(E,C)$ and clean $s(E,B)$.
It detects
\AlignedLabelFamilyFires{}/\AlignedLabelMaterialDen{} material interventions
with the conformal family rule and
\AlignedLabelUnionFires{}/\AlignedLabelMaterialDen{} with the uncorrected union.
Among \AlignedLabelSeparableDen{} aggregate-separable rows, responses are
\AlignedLabelSeparableFamilyFires{} conformal and
\AlignedLabelSeparableUnionFires{} union; all \AlignedLabelBlindDen{}
aggregate-blind aligned rows exactly equal their paired-clean sensor and rule
responses, so attack-only increment is zero; this is the structural result.
The original frozen benchmark geometry $s(E,T_y(E))$, retained for release
reproduction and comparison and structurally valid for invariance, yields
\BatchFamilyLabelFires{}/\NMaterialLabel{} conformal and
\BatchUnionLabelFires{}/\NMaterialLabel{} union, but is not the preferred
estimate of statistical response under the clean-resample geometry. Nonresponse on
separable rows is a finite-batch miss, not structural blindness. A trusted same-batch aggregate
detects all \NMaterialLabel{} exactly (Corollary~3a) and \WitnessWNine{} of
the 3,600 label rows in total; the \WitnessWTen{} relabelings that preserve
every aggregate (C1) are exposed only by the item-aligned label reference
(Corollary~3b). Feature-side coverage is mechanism-dependent: mean shift and
scaling are highly responsive because in-place perturbation smears tight
projected-feature clusters that clean resampling reproduces; the same fragile
fingerprint responds on 41--62\% of near-null controls. Feature
dropout with imputation is nearly invisible at the batch level (0--3\%) yet
changes predictions in 49--76\% of cells. Within this fixed intervention grid,
the conformal rule reduces detection only slightly:
containment in $\IXF$ is \ContainUnionIXF{} under the union and
\ContainFamilyIXF{} under the conformal rule at the primary threshold,
\ContainFiveUnionIXF{} versus \ContainFiveFamilyIXF{} at $\tau=0.05$.

Sensor dependence is substantial and explicitly quantified by the existing
frozen-only ablation. After removing KS and recalibrating the remaining family,
the core retains 3,521/4,361 $\IX$, 4,008/4,462 $\IXF$ and 3,959/4,419
$\IXFY$ primary fires, whereas the near-null retained counts fall to 10/650,
62/603 and 62/576. Thus KS dominates much of the near-null response associated
with point masses and near-duplicate clusters, while much of the core
feature-side response remains. KS stays because it was prespecified, and this
descriptive ablation does not replace the primary family.

\subsection{Offline End-to-End Decisions}

Table~\ref{tab:policy} and Fig.~\ref{fig:policy}(b,c) report the offline
end-to-end evaluation. All aggregate served fractions summarize the
prespecified equal-weight intervention grid; they are not deployment
prevalence or expected operational unsafe rates. Under P0 every one of the \NMaterial{} material
observations is served. Under the batch-level regimes the conformal-rule policy P2
serves \UnsafeFamilyIX{} ($\IX$), \UnsafeFamilyIXF{} ($\IXF$) and
\UnsafeFamilyIXFY{} ($\IXFY$) of them with clean false-action rates
\DecFprFamilyIX{}, \DecFprFamilyIXF{} and \DecFprFamilyIXFY{}, whereas the
uncalibrated union serves \UnsafeUnionIX{}, \UnsafeUnionIXF{} and
\UnsafeUnionIXFY{} at \DecFprUnionIX{}, \DecFprUnionIXF{} and
\DecFprUnionIXFY{}: the union buys \ContainUnionIXFY{} versus
\ContainFamilyIXFY{} containment in $\IXFY$ at more than four times the clean
false-action rate and \BenignHBUnionIXFY{} versus \BenignHBFamilyIXFY{}
interruption of the near-null controls. The asymmetric comparison rule served
\UnsafeLegacyFamilyIX{}, \UnsafeLegacyFamilyIXF{} and
\UnsafeLegacyFamilyIXFY{}; the adopted rule moves the count by less than 1\% of
the material rows. The served material corruptions are dominated by structure, not by
thresholds: all \NMaterialLabel{} material label observations are served
under $\IX$ and $\IXF$ (residual blind region), \UnsafeFamilyLabelIXFY{} of
them under batch-level $\IXFY$ (missed by the executed finite-batch rule), and
\UnsafeFamilyDropoutIXF{} of \NMaterialFamDropout{} material dropout
observations under $\IXF$. The $\IYm$ view provides zero containment of
material interventions in this suite; \ResidualBlindFamilyIYm{} material
cases remain in its residual blind region at $\tau\to0^{+}$, while all
\UnsafeFamilyIYm{} material cases are served once both blind and
statistically missed cases are counted.

Containment \ContainFamilyIXFYtrusted{} follows constructively from
Corollary~3 under the declared trusted references; the experiment quantifies
interruption within the prespecified near-null stress controls. Consistently, the trusted-reference regime
serves none of the \NMaterial{} material observations with zero clean false
actions on its 1,200 exact-zero rows and blocks every
non-identity relabeling and every prediction change through its item-aligned
references. Its stress-control interruption has two parts. Of the 1,200 near-null synthetic
controls the trusted conformal-rule policy interrupts
\BenignInterruptFamilyIXFYtrusted{}: \BenignHoldFamilyIXFYtrusted{} are
statistical holds by its batch component, and \TrustedGrossExactBlocks{} are
gross exact-reference blocks of controls whose predictions changed under a
0.001 perturbation. Of those exact blocks, \TrustedExactOverlap{} already
belong to the \BenignInterruptFamilyIXFY{} interruptions produced by batch
$\IXFY$/P2. The net increment is therefore
\TrustedNetAdditional{}/1,200, or \TrustedNetAdditionalPct{} percentage
points, not the gross \TrustedGrossExactBlocks{}/1,200. The controls are synthetic near-null variation,
not operational traffic, so neither number is a production false-positive
rate, and the two denominators of Table~\ref{tab:policy} are reported side by
side. The sensor-coverage-complete P3 (relative to the declared evidence
dimensions)
makes information-set conditionality literal: in $\IX$, $\IXF$ and $\IYm$
the label boundary has no coverage, so nothing is served and every clean and
near-null row is interrupted; in $\IXFY$ and $\IXFYs$ every boundary is
covered and P3 coincides with P2, including P2's \UnsafeFamilyIXFY{}
materially altered audit results served in $\IXFY$: abstention on missing coverage is not a guarantee of
power, and the choice between P2 and P3 is a declared risk decision, not a
detector property. The composition with the six frozen contract envelopes
reproduces their actions (supplement).

\begin{table}[!t]
\caption{Offline decisions on 10,800 unchanged frozen interventions (7,008 conclusion-changing at the structural-sensitivity endpoint $\tau\to0^{+}$; $\tau=0.05$ is in the supplement). P1 is union, P2 conformal and P3 abstains on missing sensor coverage. Batch FPR uses 12,000 clean draws from evaluation pools disjoint from calibration pools; draws within a pool may overlap. The starred exact-zero check uses 1,200 rows and is a different estimand. $^{\ddagger}$P3 holds every row of a regime whose mandatory label or prediction boundary has no declared coverage; that cell is abstention by construction, not a detector false-action rate. Benign is interruption of 1,200 near-null controls; served counts are altered audit results, not operational events.}
\label{tab:policy}
\centering\scriptsize\setlength{\tabcolsep}{2pt}
\begin{tabular}{@{}llrrrrr@{}}
\toprule
Regime & Policy & FPR/check & Benign & Served & Contain. & Blind \\
\midrule
$\mathcal I_X$ & P1 union & 0.123 & 0.75 & 4,429 & 0.37 & 2,617 \\
$\mathcal I_X$ & P2 conf. & 0.056 & 0.54 & 4,496 & 0.36 & 2,617 \\
$\mathcal I_X$ & P3 abst. & n/a$^{\ddagger}$ & 1.00 & 0 & 1.00 & 2,617 \\
\addlinespace[1pt]
$\mathcal I_{XF}$ & P1 union & 0.201 & 0.75 & 4,233 & 0.40 & 2,617 \\
$\mathcal I_{XF}$ & P2 conf. & 0.059 & 0.50 & 4,368 & 0.38 & 2,617 \\
$\mathcal I_{XF}$ & P3 abst. & n/a$^{\ddagger}$ & 1.00 & 0 & 1.00 & 2,617 \\
\addlinespace[1pt]
$\mathcal I_{Y_m}$ & P1 union & 0.048 & 0.00 & 7,008 & 0.00 & 5,450 \\
$\mathcal I_{Y_m}$ & P2 conf. & 0.048 & 0.00 & 7,008 & 0.00 & 5,450 \\
$\mathcal I_{Y_m}$ & P3 abst. & n/a$^{\ddagger}$ & 1.00 & 0 & 1.00 & 5,450 \\
\addlinespace[1pt]
$\mathcal I_{XFY}$ & P1 union & 0.257 & 0.75 & 4,182 & 0.40 & 0 \\
$\mathcal I_{XFY}$ & P2 conf. & 0.055 & 0.48 & 4,394 & 0.37 & 0 \\
\addlinespace[1pt]
$\mathcal I_{XFY}^{\star}$ & P2/P3 (exact) & 0.000 & 0.51 & 0 & 1.00 & 0 \\
\bottomrule
\end{tabular}
\end{table}

\subsection{Adaptive Cluster-Preserving Attacker}

\begin{figure}[!t]
  \centering
  \includegraphics[width=\columnwidth]{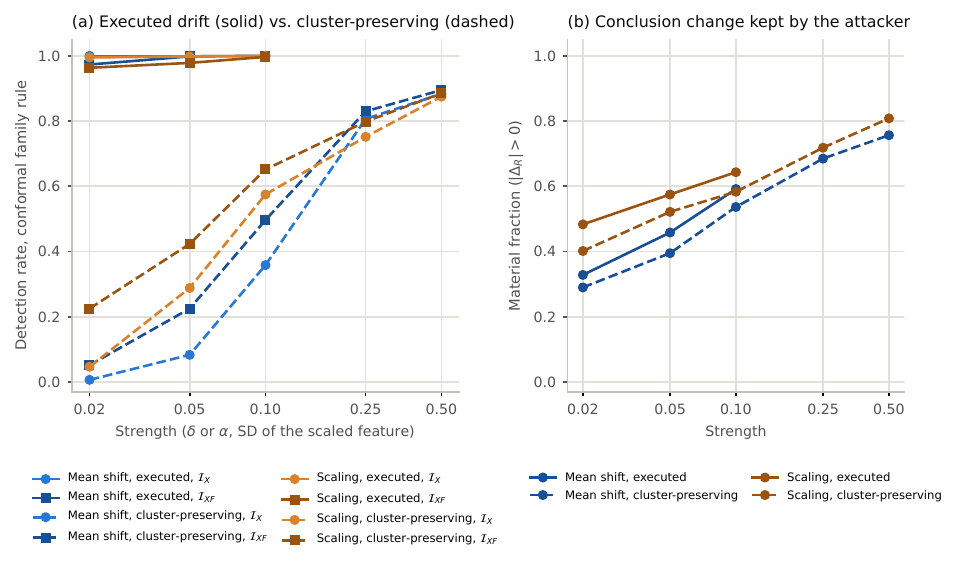}
  \caption{Corrected-JSD response on the original frozen Gate~A rows. (a) Detection under the conformal family rule of the
  executed drift mechanisms (solid) and of their cluster-preserving variants
  (dashed) in $\IX$ and $\IXF$, 600 observations per point. (b) Material
  fraction of the same rows: the construction reduces response against the
  declared batch-level fingerprint at matched strengths while keeping most of
  the conclusion changes.}
  \label{fig:adversarial}
\end{figure}

The frozen Gate~A stress test used clean-resample calibration but scored attacked
batches as $s(E,T(E))$. Its matched response falls from
\AdvDetCtrlHeadlineMin--\AdvDetCtrlHeadlineMax{} to
\AdvDetAdaptHeadlineMin--\AdvDetAdaptHeadlineMax{} while retaining
\AdvMatRetentionMinPct--\AdvMatRetentionMaxPct\% materiality; this is
design-specific and against the declared cluster-sensitive fingerprint, not
evidence of general evasion of drift detectors. Dataset cluster structure
contributes to the response, and generalization beyond this fingerprint and
these staged datasets remains open.
Figure~\ref{fig:adversarial} shows the matched original-geometry response and
materiality profiles used for this declared-fingerprint stress test.

The aligned sensitivity confirms the qualitative ordering. Exact paired
identity is \GeoIdentityMax{}, distinct from the clean-resample false-action
rate (family rates \GeoCleanFamilyIX--\GeoCleanFamilyIXF). Mean shift and scaling remain
almost fully responsive; dropout is \GeoDropoutFamilyMin--\GeoDropoutFamilyMax{}
and near-null response spans \GeoTinyGaussianIXF--\GeoTinyScalingIX. Adaptive
response is lower in \GeoGateLowerCellsMin--\GeoGateLowerCellsMax{} of
\GeoGateCellTotal{} environment/split cells, never higher, with materiality of
\GeoGateAdaptiveMaterialMin--\GeoGateAdaptiveMaterialMax{}. Full
strength/regime results are in the supplement. Same-item scoring amplified part
of the original contrast, but aligned Gate~A still shows that calibration
controls clean-resample false actions without guaranteeing attack coverage.

\subsection{Quantum-Workflow Coverage}

All 165 dependent design cells (11 conditions $\times$ five splits $\times$ three
dimensions) and all nine quantum-gate acceptance checks pass. In this ideal-statevector
gate one anchored comparison against $\Phi(C_0)$ is a semantic probe on
circuit-side rows and an observed-kernel reference on post-processing rows;
finite-shot estimation separates those anchors. Approved rewrites change
provenance but preserve semantics (class (a)); PSD-preserving substitution
escapes algebra and circuit provenance but not the anchored
$K_{\mathrm{obs}}$ comparison (class (c)); both shot emulators yield non-zero
repeated-estimation discrepancy in their 30 cells without establishing a
universal rate (class (b)). Full per-condition responses, contract decisions
and the secondary model profile are in the supplement. This is a bounded
simulator instantiation of Proposition~7, not QPU/provider validation.

\section{Integrity-Audit Contract}

Each contract names its boundary, intervention class, view/reference, premise,
residual region and response. For labels, $R_0$, $M_0$ and item alignment bind
conclusion, aggregate and identity claims. The quantum contract combines
provenance, approved equivalence, anchored kernels, algebra and repeated
estimation. Exact-invariant violations block; missing mandatory evidence holds;
P2 may serve within a residual region, while P3 promises no minimum power.

\section{Discussion and Limitations}

\subsection{Security Interpretation}

A metric can change while predictor output is unchanged: input robustness
cannot repair corrupted evaluation labels. Zero response establishes blindness
only when the evidence could not distinguish the states; otherwise power
depends on the null. The 27--39\% served fractions summarize the prespecified
Gate~A adaptive cluster-preserving intervention grid, not deployment prevalence.
Gate~A is a stress test of one cluster-sensitive fingerprint; its aligned sensitivity preserves
the control/adaptive ordering while exposing a geometry effect.

Evidence sufficiency and necessity are claim-relative within the declared
reference lattice, where granularity means authenticated or bound information,
not digest length. A Merkle root or
position-binding vector commitment can compactly bind level-C item identity
\cite{Merkle1988,CatalanoFiore2013}. Level B instead lets aggregate- or
conclusion-only audits verify $M_0$ or $R_0$ without retaining or exposing the
complete item-wise label store, which can reduce disclosure and support
separation of duties over sensitive labels. This is evidence minimization, not
a new cryptographic primitive or legal guarantee. An external root remains
necessary (Proposition~6); this is neither a deployed protocol nor a minimum
over every possible audit architecture.

\subsection{Validity Boundaries}

The eight fixed settings are not sampled deployments, and balanced staging
does not reproduce prevalence. The quantum branch uses ideal-statevector and
finite-shot emulation, not QPU/provider validation. The secondary SVC/QSVC
profile does not isolate kernel geometry. Structural blind regions are
independent of the number of rows: a view-preserving intervention remains
invisible as $n$ grows. For interventions that do change an observed statistic,
however, power can increase with batch size as sampling variability often
decreases, approximately as $n^{-1/2}$ for proportion-like statistics under
regular conditions. The primary geometry-aligned outcomes---
\AlignedLabelFamilyFires{}/\AlignedLabelMaterialDen{} conformal and
\AlignedLabelUnionFires{}/\AlignedLabelMaterialDen{} union---and the original
frozen outcomes---\BatchFamilyLabelFires{}/\NMaterialLabel{} conformal and
\BatchUnionLabelFires{}/\NMaterialLabel{} union---are finite-design results,
not power bounds. E1 is the
only 256-row environment and exhibits non-zero label response, but because
batch size is confounded with environment and null behavior, this cannot
identify a batch-size effect.

The conformal level requires exchangeability; overlapping draws, different
pool rows and E1 violate it, so executed rates are descriptive. Gate~A fixes
one fingerprint and two mechanisms; generalization beyond its dataset
structure remains open. The algebraic quantum check concerns the raw
estimated/observed kernel before downstream PSD repair; a repaired-kernel
consumer defines a different, unevaluated observable path.

The trusted item-aligned reference is an assumption under Proposition~6. P3
abstains on missing coverage, not low power. All policies are offline; runtime
non-stationarity, recovery, QPU execution and service enforcement remain open.

\section{Conclusion}

Hybrid workflow integrity is evidence-relative. Interventions can be
structurally invisible, statistically missed in a finite batch, or show reduced
response against the declared monitored fingerprint. The geometry-aligned
label construction is the primary statistical interpretation; its 764
aggregate-blind rows retain zero attack-only increment. The substantial KS
dependence is quantified without replacing the prespecified family. The
conformal guarantee remains conditional on exchangeability and the quantum
instantiation simulator-bounded. Within the declared lattice, authenticated
$R_0$, $M_0$ and item alignment certify conclusion,
aggregate-plus-conclusion and item-identity claims, respectively. Assurance
must state its view, trust root, premise, residual region and response.

\section*{Data, Code, and Ethics Statement}

The version 1.3.9 artifact contains code, tests, frozen and corrective derived evidence and eleven manifests;
raw benchmarks are not redistributed. Sources, hashes and staging are
documented. Its version DOI is 10.5281/zenodo.22750616 under concept DOI
10.5281/zenodo.22550852; immutable predecessor v1.3.8 has DOI
10.5281/zenodo.22706167.
Code is Apache-2.0 and derived evidence/documentation
CC~BY~4.0. No new human-participant or personal-data collection was performed.

\section*{Acknowledgment}

OpenAI Codex and Anthropic Claude Code assisted with drafting and language
review in the Abstract, Introduction, Related Work, Methods, Results,
Discussion, data/code statement and supplement; targeted literature-search
support; code generation and inspection for analysis, tests and release tools;
and table, figure, artifact and submission-package preparation. Human authors
independently validated every formal statement and proof, reference, code path,
experiment, result and final passage and retain full responsibility. Neither
system is an author or proof authority; detailed disclosure accompanies the
submission.

\bibliographystyle{IEEEtran}
{\sloppy\bibliography{references}}

\end{document}